\documentclass[conference]{IEEEtran}
\IEEEoverridecommandlockouts

\usepackage{cite}
\usepackage{amsmath,graphicx, cite, booktabs, subfig}
\usepackage{amsmath,amssymb,amsfonts,hyperref
}
\usepackage{algorithmic}
\usepackage{graphicx}
\usepackage{textcomp}

\usepackage{xcolor}
\def\BibTeX{{\rm B\kern-.05em{\sc i\kern-.025em b}\kern-.08em
    T\kern-.1667em\lower.7ex\hbox{E}\kern-.125emX}}
\begin{document}

\title{Latent Watermarking of Audio Generative Models}

\author{
\IEEEauthorblockN{Robin San Roman}
\IEEEauthorblockA{\textit{Meta, FAIR} \\
\textit{Inria Nancy}}
\and
\IEEEauthorblockN{Pierre Fernandez}
\IEEEauthorblockA{\textit{Meta, FAIR}\\ 
\textit{Inria Rennes}}
\and
\IEEEauthorblockN{Antoine Deleforge}
\IEEEauthorblockA{\textit{Inria Nancy
}}
\and
\IEEEauthorblockN{Yossi Adi}
\IEEEauthorblockA{\textit{Meta, FAIR} \\
\textit{Hebrew University of Jerusalem}}
\and
\IEEEauthorblockN{Romain Serizel}
\IEEEauthorblockA{\textit{Inria Nancy}}

}

\newcommand{\modelname}{StemLM }
\maketitle

\begin{abstract}
The advancements in audio generative models have opened up new challenges in their responsible disclosure and the detection of their misuse. In response, we introduce a method to watermark latent generative models by a specific watermarking of their training data.
The resulting watermarked models produce latent representations whose decoded outputs are detected with high confidence, regardless of the decoding method used. This approach enables the detection of the generated content without the need for a post-hoc watermarking step.
It provides a more secure solution for open-sourced models and facilitates the identification of derivative works that fine-tune or use these models without adhering to their license terms.
Our results indicate for instance that generated outputs are detected with an accuracy of more than 75\% at a false positive rate of 10\textsuperscript{-3}, even after fine-tuning the latent generative model. \vspace*{3pt}
\end{abstract}


\section{Introduction}

Sophisticated generative models are impacting various audio modalities: environmental sounds~\cite{kreuk2022audiogen, audioldm}, music~\cite{musiclm, copet2024simple}, and speech~\cite{valle, audiobox, audiolm}. 
These models produce outputs increasingly indistinguishable from real data~\cite{stablediff, gpt4}. Their rapid proliferation and quality raise concerns about misuse (e.g. creation of deepfakes) and respect for intellectual property. 
These concerns are heightened when models are open-sourced, since they can be easily accessed and used by anyone, including malicious actors.
Consequently, regulators suggest watermarking to label and detect generative model outputs (refer to the \href{https://artificialintelligenceact.eu/}{EU AI Act}, \href{https://www.whitehouse.gov/briefing-room/presidential-actions/2023/10/30/executive-order-on-the-safe-secure-and-trustworthy-development-and-use-of-artificial-intelligence/}{White House executive order}, and \href{http://www.cac.gov.cn/2023-07/13/c_1690898327029107.htm}{CAC measures}).

Watermarking is a technique that slightly alters the audio after its generation, in a way that is inaudible for humans but identifiable by specific detection algorithms.
The state-of-the-art methods are based on deep neural networks~\cite{audioseal, wavmark} that are trained end-to-end to embed and detect watermarks in audio signals, even after audio compression or editing.
Such methods are for instance employed to safeguard APIs for public model demonstrations~\cite{audiobox, communication2023seamless}.
However, while post-hoc watermarking has proven effective in certain scenarios, it is not as effective for protecting open-sourced models, as malicious users could potentially extract the output before the watermarking stage (for example by commenting out the code responsible for watermark embedding).

In the image domain, some methods~\cite{stablesignature, kim2023wouaf} fine-tune decoders to output watermarked images directly, to make it compliant with open-sourcing.
However, in the audio domain, it is common and cost-effective to train decoders (also called vocoders) that convert latent representations to waveforms~\cite{kumar2019melgan, kong2020hifi, mbd}.
Watermarking can thus be easily bypassed by using non-watermarked vocoders. 
Therefore, in this article, we propose to watermark the latent generative model that creates the latent representations.

We focus on MusicGen~\cite{copet2024simple} due to its performance and adoption.
It consists of an auto-encoder EnCodec~\cite{encodec} that compresses audio into discrete representations (tokens) and a single-stage transformer (audio Language Model, LM) that predicts the next tokens and decodes them into a music stream.
We train watermark generator/detector models to be robust to EnCodec.
Intuitively, this makes both the audio and encoded tokens watermarked.
We then train the LM on tokens derived from audios that were preemptively marked.
The resulting LM produces tokens whose decoded outputs are watermarked, irrespective of the LM conditioning or decoding algorithm.
In other terms, as long as the watermarking algorithm withstands the audio tokenization, the watermark transfers from the training data to the generative model outputs.

In short, 
(1) we introduce a way to watermark audio generative models at the latent representations level, 
(2) we demonstrate that it makes generations detectable with high confidence while having almost no influence on the model performance, 
(3) we demonstrate the robustness of the watermark to model-level changes, namely, switching the decoding algorithm and fine-tuning the audio LM.

\begin{figure*}[th!]
    \centering
    \includegraphics[width=1.0\linewidth]{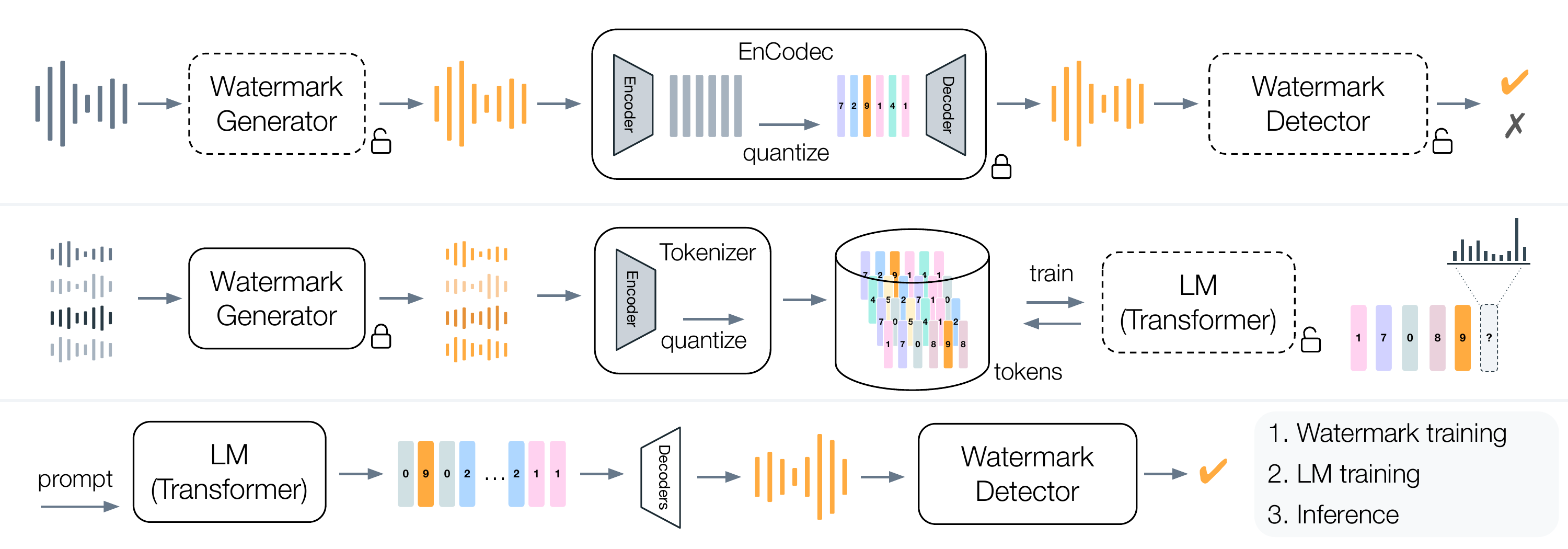}
    \caption{
        \textbf{Overview of our method.} (1.) We train a watermark generator and detector based on AudioSeal~\cite{audioseal}, enhancing robustness against EnCodec~\cite{encodec} by processing the watermarked audio through EnCodec before detection. (2.) We watermark the audios from our database and train a MusicGen~\cite{copet2024simple} model for next token prediction on this watermarked data. (3.) During inference, we prompt (using text or audio) the language model and decode audios that are detectable with watermark detector.
    }
    \label{fig:fig1}
\end{figure*}

\section{Related Work}

\textbf{Audio generation} is a challenging task because audio signals are high-dimensional and have complex temporal dependencies.
Early autoregressive deep-learning-based approaches like WaveNet~\cite{oord2016wavenet} were quickly followed by GAN-based models~\cite{kumar2019melgan, kong2020hifi}.
Inspired by progress in text generation~\cite{radford2018improving, brown2020language}, audio language models (LM) have recently emerged as state of the art for most audio generative tasks such as text-to-speech~\cite{kharitonov2021text, valle, communication2023seamless}, music~\cite{copet2024simple, musiclm} or sound~\cite{kreuk2022audiogen} generation.
They make audio modeling more tractable by compressing audio into discrete tokens using models like EnCodec~\cite{encodec}, SoundStream~\cite{soundstream}, or DAC~\cite{dac}. 
Additional tokens coming from text, melody, phoneme, speaker embedding, etc. may serve as conditioning to generate audio with user-specific characteristics.
Then a transformer-based model~\cite{vaswani2017attention} generates audio by predicting the next tokens and decoding them.

In parallel to audio LMs, latent diffusion models have also been largely studied in recent works on audio generation. Those models can sample in an non autoregressive way from the training data distribution and have recently shown great generative habilities on different audio modalities such as speech~\cite{voicebox,audiobox, naturalspeech2}, music~\cite{jasco, stableaudio} or general audio~\cite{audioldm}.

\vspace{0.5em}\noindent
\textbf{Invisible audio watermarking}
has evolved from using domain-specific features in the time/frequency domain of audios~\cite{lie2006robust, kalantari2009robust} to deep learning methods that employ encoder/decoder architectures~\cite{wavmark, liu2023dear, pavlovic2022robust, o2024maskmark}. 
Notably, AudioSeal~\cite{audioseal} introduces localized audio watermarking with a detector producing time-step-level logits. This method also allows for a watermarking robust to neural compression models which is a necessary element for our work.

\vspace{0.5em}\noindent
\textbf{Generative model watermarking} is attracting renewed interest thanks to its potential to improve detection of AI-generated contents.
In this context, the aforementioned methods apply watermarking \textit{post-hoc} after audio generation, unlike more recent methods which do so \textit{in-model}.
Examples include watermarking: image GANs by training a hyper-network model~\cite{yu2021responsible}, latent diffusion models by a quick fine-tuning of their decoder~\cite{stablesignature}, and HiFi-GAN decoders that take mel-spectrograms and output waveforms~\cite{juvela2023collaborative}. 
Unlike the last two approaches, our method operates one step earlier at the latent representation level.
It draws inspiration from research demonstrating that watermarks embedded in images or texts may propagate from the training data of generative models to their outputs~\cite{yu2021artificial, zhao2023recipe, gu2023learnability, sander2024watermarking}. 
We apply this concept to audio generative models and target audio language models.

\section{Audio Model Watermarking}

\subsection{Problem statement}

We consider providers training an audio-generative model on a large proprietary dataset. 
They aim to release the model publicly but worry about misuse and unauthorized redistribution. 
To mitigate these concerns, they watermark the model during training to enhance the detection of generated content or unauthorized API usage. 
We describe this watermarking process below and provide a step by step overview in Figure~\ref{fig:fig1}.

\subsection{Audio watermarking}\label{sec:audio-watermarking}

We first build an audio watermarking model based on AudioSeal~\cite{audioseal}.
It jointly trains a watermark generator $G$ and a watermark detector $D$. 
$G$ takes a signal $s \in \mathbb{R}^T$ and generates an additive watermark $\delta_w$, that is made imperceptible through perceptual losses $\mathcal{L}_{\textrm{percep}}(s, s+\delta_w)$.
The watermarked audio $s+\delta_w$ is augmented into $s'$. 
Augmentations include padding the audio with 0, replacing intervals of watermarked audio with non-watermarked audio from the same batch, or dropping the $\delta_w$.
$s'$ is fed to the detector, which is trained to output which part of a waveform is watermarked via a localization loss $\mathcal{L}_{\textrm{loc}}(D(s'), y')$, 
where $y' \in \{0,1\}^T$ indicates the watermarked intervals in $s'$ ($1$ for watermarked, $0$ otherwise).

We make the following changes with regard to AudioSeal's recipe.
We remove the message encoder to only focus on watermark detection.
Furthermore, it is important to remember that the audio LM will be trained on tokens, not directly on audio. 
Hence, the LM will not retain watermark information if it is absent at the discrete representation level. 
Therefore, we train the watermark generation/detection to be very robust to the specific EnCodec compression model later used for audio tokenization (see Sec.~\ref{sec:tm-training}).
This is done by oversampling this EnCodec augmentation so that 50\% of batches go through EnCodec before the detection phase. %

\subsection{Audio language model}\label{sec:tm-training}

We select MusicGen~\cite{copet2024simple} as the audio LM to watermark. 

\vspace{0.5em}\noindent
\textbf{Watermarking.}
The first step is to watermark the audios with the model of Sec.~\ref{sec:audio-watermarking}.
This is done on the fly at loading time (this takes around $10$ ms for a $10$-second audio).

\vspace{0.5em}\noindent
\textbf{Tokenization.}
We use the EnCodec compression model to transform the audio signal into a discrete sequence of tokens suitable for language modeling.
It uses residual vector quantization (RVQ)~\cite{soundstream} which compresses an audio signal $s \in \mathbb{R}^T$ into $K$ streams of tokens $(u^{(j)}_i ) _{j \in [1,K] \, ;\, i \in [0, T / f_r]}$ ($f_r$ being the frame rate).
This model is trained on audio segments sampled at a rate of $32$~kHz and $f_r = 50$~Hz, the number of codebooks is $K=4$ and the codebook size is $2048$ ($u^{(j)}_i \in [1,2048]$). Overall, this results in a overall bit rate of $2.2$~kbit per second

\vspace{0.5em}\noindent
\textbf{Language modeling} aims to build a probabilistic model of sequences of discrete tokens.
MusicGen implements a delay pattern~\cite{kharitonov2021text} that adds a delay $k$ to the $k$-th residual. 
It allows the model to generate the tokens in a coarse to fine order. 
This way all the streams can be sampled in parallel while assuring that all the previous residuals are fixed when sampling a given token.
Put differently, the transformer is fed with a sequence of embeddings created from $K$ tokens: $s_i = \{u^{(4)}_{i-4}, u^{(3)}_{i-2}, u^{(2)}_{i-1}, u^{(1)}_{i} \}$.
The embedding of $s_i$ is then the sum of the embedding of each of its constitutive tokens, with additional sinusoidal embeddings. 
As most current language models~\cite{gpt4,touvron2023llama}, training is done with next token prediction and the cross-entropy is computed per codebook.

\section{Experiments}
\subsection{Experimental details}

\textbf{Watermarking models.} 
We train on an internal music dataset containing 1.5k songs at $32$~kHz sample rate, with 1-second audio excerpts.
The model is trained for $400$k steps with batch size $64$. Hyper-parameters (network architectures, optimizer...) are kept the same as in the original work~\cite{audioseal}.

\vspace{0.5em}\noindent
\textbf{MusicGen models.} 
We use 20k hours of licensed music to train the models with two different model sizes: small ($300$M parameters) and medium ($1.5$B parameters).
They are similar in quality and diversity to the ones used by Copet et al.~\cite{copet2024simple}. 
We use public implementation and default parameters available on the \href{https://github.com/facebookresearch/audiocraft}{AudioCraft GitHub page}.
Each model is trained for $200$ epochs with a batch size of $128$, with the default optimizer.
We use $64$ GPUs for the medium model and $32$ for the small.

\vspace{0.5em}\noindent
\textbf{Inference.} 
For music generation sampling, we use top-k sampling with $k=250$ tokens and a temperature of $1.0$.

\subsection{Quality of the audio generative model}

We first subjectively evaluate how watermarking influences the quality of the generative models.

We adhere to the original paper's protocol~\cite{copet2024simple}, using (OVL) to assess sample quality and (REL) to evaluate relevance to text prompts.
Models are tested on 15-second generation using 40 text prompts from the test set of MusicCaps~\cite{musiclm}. 
Every sample is rated by $20$ listeners that rate them on a scale from 1 to 100. 
For every study, we report both mean score and CI95. 
\autoref{tab:mos} shows that the performance difference between a model trained on watermarked data and one trained on normal data is negligible.
This holds true for both sizes, with the rating difference falling within the confidence interval.

\begin{table}[t]
\centering
\caption{
    \textbf{Subjective evaluation.} 
    We compare audio quality and text relevance of the original MusicGen models (ori.) and our models that natively outputs watermarked audios (ours).
}\label{tab:mos}
\begin{tabular}{lll}
    \toprule
    Size   & \textsc{Ovl} ($\uparrow$)  & \textsc{Rel} ($\uparrow$) 
    \\
    \midrule
    Ground Truth & $93.08 \pm 0.53$ & $93.01 \pm 0.68$ \\
    \midrule
    
    Small (ori.)  & $83.67 \pm 1.85 $ & $82.42
     \pm 1.37$ \\
    Small (ours) & $\textbf{84.13} \pm 2.21$ & $\textbf{82.46}  \pm1.44
     $    \\
    \midrule
    Medium (ori.)  & $ \textbf{85.92} \pm 1.46$ & $\textbf{83.71} \pm 1.79$ \\
    Medium (ours) &   $84.91 \pm 1.53$    &   $ 82.64 \pm1.41 $      
    \\
    \bottomrule
\end{tabular}

\end{table}

\subsection{Detection and localization results}\label{sec:detection}

\textbf{Detection.}
To evaluate the detection performance, we generate 10k positive 15-second samples with the watermarked model and use 10k negative samples from our test set that we compress using the codec model. 
The watermark detector gives a score per time-step of the audio, which we average to get a score for the whole audio.
Audio is flagged as watermarked if this score is higher than a threshold $\tau$.
We report in Tab.~\ref{tab:detect} the area under the ROC curve, as well as the accuracy for the best $\tau$ (and true positive and false positive rates at this $\tau$, TPR, and FPR).
The generated output is indeed watermarked as indicated by the detection metrics: the AUC is close to 1 and TPR is higher than $0.95$ at FPR around $10^{-4}$.

\begin{table}[t]
\centering
\caption{
     \textbf{Detection and localization results} on 10k positive/negative samples, for the MusicGen models watermarked post-hoc using AudioSeal (+AS), or in-model with our method (ours).
    We report the area under the ROC curve (AUC), the accuracy for the best threshold, as well as the intersection over union (IoU) and sample level accuracy (SL-Acc.).
    }
    \label{tab:detect}
    \resizebox{1.0\linewidth}{!}{
        \begin{tabular}{l *{3}{l}  *{2}{c}}
        \toprule
        & \multicolumn{3}{c}{Detection} & \multicolumn{2}{c}{Localization} \\
        \cmidrule(rr){2-4} \cmidrule(rr){5-6}
        Model   &  AUC & Acc.   & TPR / FPR    & IoU & SL-Acc.
        \\
        \midrule
        Small + AS  & 1.0 & 1.0 & 1.0 / $0.0$ & 1.0 & 1.0 \\
        Medium + AS & 1.0 & 1.0 & 1.0 / $0.0$ & 1.0 & 1.0 \\
        \midrule
        Small (ours)  & 0.999 & 0.993 & 0.986 / $2.10^{-4}$ & 0.81 & 0.91 \\
        Medium (ours) & 0.999 & 0.994 & 0.988 / $3.10^{-4}$ & 0.91 & 0.96 \\
        \bottomrule
        \end{tabular}
    }

\end{table}

\vspace{0.5em}\noindent
\textbf{Localization.}
We then evaluate if the detector still has the property to locally detect watermarked segments.
To do so, we generate 15-second samples and replace parts with other non-watermarked audio from our test set. The proportion of signal that is watermarked is 50\% on average.  We then measure the precision of the detection using the detection accuracy at the sample level together with the intersection over union (IoU) metric. For localization, we use a fixed detection threshold set at $0.5$.
As shown in Tab.~\ref{tab:detect}, results are on-par (although a bit lower) to AudioSeal, showing that the detector keeps a good-enough performance on the localization of generated outputs.

\vspace{0.5em}\noindent
\textbf{Robustness.} We evaluate the robustness to different audio edits, and compare the performance of our in-model watermarking and of post-hoc watermarking that directly applies the watermark to generated outputs with AudioSeal. 
The evaluation is made on 10k samples.

\newcommand{\aux}[1]{{\scriptsize{\textcolor{gray}{#1}}}}
\begin{table}[t]
\centering
\caption{
    \textbf{Robustness}
    to audio edits for post-hoc watermarking with AudioSeal or using our in-model method.
    The results are computed on audios generated with the Small model.
} \label{tab:robustness}
        \resizebox{1.\linewidth}{!}{
        \begin{tabular}{lllll}
        
        \toprule
        & \multicolumn{2}{c}{Post-hoc} & \multicolumn{2}{c}{In-model} \\
       \cmidrule(rr){2-3} \cmidrule(rr){4-5}
        Edit   &  AUC & Acc \aux{TPR / FPR} & AUC & Acc \aux{TPR / FPR}  \\
         \midrule
        White Noise & 0.995 & 0.99 \aux{0.99/0.01} & 0.941 & 0.93 \aux{0.91/0.04}\\
        Lowpass & 0.99 & 0.99 \aux{0.99/0.01} & 0.942 & 0.94 \aux{0.91/0.03}\\
        Highpass & 0.931 & 0.98 \aux{0.98/0.2} & 0.941 & 0.93 \aux{0.91/0.04}\\
        Resample & 0.940 & 0.94 \aux{0.91/0.03} & 0.999 & 0.99 \aux{0.99/0.01}\\
        AAC  & 0.999 & 0.98 \aux{0.97/0.01} & 0.88 & 0.81 \aux{0.81/0.04}\\
        Pink Noise & 0.996 & 0.97 \aux{0.98/0.03} & 0.956 & 0.92 \aux{0.91/0.06}\\
        Echo & 0.950 & 0.98 \aux{0.99/0.02} & 0.906 & 0.92 \aux{0.89/0.04}\\
        \bottomrule
        \end{tabular}}

\end{table}

\autoref{tab:robustness} shows that while in-model watermarking keeps a decent robustness to common audio edits there is a slight performance decrease compared to the original watermarking model. 
Therefore, when post-hoc watermarking is feasible, it might be preferable to in-model watermarking. 
The latter is better suited for scenarios where post-hoc watermarking is not possible, such as when open-sourcing a model.

\section{Attacks on the Model's Watermark}

We now focus on model-level attacks, \textit{i.e.}, modifications of the model attempting to make its outputs undetectable.
\subsection{Switching decoder}

Previous works alter the latent decoder to embed the watermark~\cite{stablesignature, juvela2023collaborative}. 
However, audio vocoders are relatively easy to train and interchange~\cite{kong2020hifi, kumar2019melgan, kong2021diffwave}.
They do not necessitate extensive data or computational power compared to those required for training an audio LM. 
Therefore, replacing the decoder to use the watermark-free generative model is rather straightforward. 
In contrast, our work embeds the watermark at the latent stage for robustness against decoder changes.

We now evaluate how the change of the decoder influences detection performance.
In previous experiments, the default decoder was the codec model from MusicGen. 
We replace it with Multi-Band Diffusion~\cite{mbd} which uses diffusion to map discrete EnCodec tokens into the waveform domain, and a discrete version HiFi-GAN~\cite{kong2020hifi}, which we trained on tokens-waveform pairs.
We use the same experimental setup as in Sec.~\ref{sec:detection}, but with different algorithms to decode the tokens. 
\autoref{tab:decoders} shows that changing the decoder has very little impact on the detection metrics.
Notably, using the diffusion-based decoder reduces the AUC only by around $0.01$.

\begin{table}[t]
    \centering
    \caption{
    Detection results with other decoding algorithm.
    }
    \label{tab:decoders}
    \small
    \begin{tabular}{l lll}
    \toprule
       Decoder & AUC & Acc. & TPR / FPR  \\
       \midrule
         HiFi-GAN               & 0.999 & 0.990 & 0.980 / $4.10^{-4}$ \\
         Multi-band diffusion   & 0.991 & 0.954 & 0.951 / 0.043 \\
         \midrule
         Default                & 0.999 & 0.993 & 0.986 / 0.000 \\
       \bottomrule
    \end{tabular}
\end{table}

\subsection{Model fine-tuning}

One potential attack could be to remove the watermark through ``model purification'', which involves fine-tuning the language model on a non-watermarked dataset. 
To test this we fine-tune with different learning rates the small version of the model using a different internal music dataset $D_{FT}$ of similar size without watermarks.
For each setting, the model is trained for 20 epochs (10\% of the total pre-training steps). 
We then generate 10k samples with the purified models and obtain scores through the watermark detector.

For each experiment, we report the accuracy obtained for the best threshold on the detection score as well as the TPR when the threshold is chosen such that the FPR is at $10^{-3}$. 
We also report the Fr\'echet Audio Distance (FAD)~\cite{fad} that evaluates the quality of the generative model.
We include as a reference the performance of the model before fine-tuning and of a model trained from scratch on $D_{FT}$.

\begin{table}[t]
\centering
\caption{Performances of the model after fine-tuning for 20 epochs with different learning rates on a non watermarked dataset. No FT is the model before fine-tuning, scratch is a model trained from scratch on the non watermaked dataset.}\label{tab:fine_tuning}
\resizebox{0.9\linewidth}{!}{
    \begin{tabular}{lccc}
        \toprule
        LR           & Best Acc. & TPR @ FPR=$10^{-3}$ & FAD ($\downarrow$)   \\
        \midrule
        No FT        &  0.993   & 0.983  & 2.067 \\  %
        \midrule
        $ 2\times10^{-6}$  &  0.964  &   0.765   &  2.908 \\ %
        $ 5\times10^{-6}$ &  0.928  &   0.456  & 3.149 \\ %
        $ 2\times10^{-5}$ &  0.833  &   0.079   &  3.344\\ %
        $ 5\times10^{-5}$ &  0.721  &   0.019    & 3.740 \\ %
        $ 1\times10^{-4}$ &  0.721  &   0.018  & 3.643 \\ %
        \midrule
        Scratch      &  N/A &  N/A   &  3.839 \\
        \bottomrule
    \end{tabular}
}
\end{table}

\autoref{tab:fine_tuning} suggests that a higher learning rate during fine-tuning makes watermarks more difficult to detect, but it also causes the distribution of generated data to deviate further from the protected model's dataset; at larger learning rates, the distribution has a similar FAD to a model trained from scratch on different data.
In other words, since the FAD is almost the same as a model trained from scratch, it may not be worthwhile to start from the watermarked model.

\section{Conclusion \& Discussion} 

This work introduced a straightforward yet effective approach to watermarking audio language models.
This is done via watermarking their training data in a way that is robust to the compression algorithm used to create tokens.
It does not require modifications to the model architecture or the training process.
Our method is the first to watermark at the latent level and is robust to changes in the decoding process.
The main drawback of the current approach is that it requires training the model from scratch, which may be difficult for versioning large models or for adapting already-trained models.
While there is a slight decrease in robustness to audio edits compared to post-hoc watermarking, this method allows to keep the watermark in situations for which post-hoc watermarking is not suitable (open sourcing...). 
In conclusion, watermarking can help trace content origin and support regulatory efforts. 
It is not a standalone solution and should be complemented with measures like policies, education, or monitoring.

\bibliographystyle{IEEEbib}
\bibliography{mybib}

\begin{thebibliography}{10}

\bibitem{kreuk2022audiogen}
Felix Kreuk, Gabriel Synnaeve, Adam Polyak, Uriel Singer, Alexandre D{\'e}fossez, Jade Copet, Devi Parikh, Yaniv Taigman, and Yossi Adi,
\newblock ``Audiogen: Textually guided audio generation,''
\newblock {\em arXiv preprint arXiv:2209.15352}, 2022.

\bibitem{audioldm}
Haohe Liu, Zehua Chen, Yi~Yuan, Xinhao Mei, Xubo Liu, Danilo Mandic, Wenwu Wang, and Mark~D Plumbley,
\newblock ``{AudioLDM}: Text-to-audio generation with latent diffusion models,''
\newblock {\em Proceedings of the International Conference on Machine Learning}, 2023.

\bibitem{musiclm}
Andrea Agostinelli, Timo~I. Denk, Zalán Borsos, Jesse Engel, Mauro Verzetti, Antoine Caillon, Qingqing Huang, Aren Jansen, Adam Roberts, Marco Tagliasacchi, Matt Sharifi, Neil Zeghidour, and Christian Frank,
\newblock ``Musiclm: Generating music from text,'' 2023.

\bibitem{copet2024simple}
Jade Copet, Felix Kreuk, Itai Gat, Tal Remez, David Kant, Gabriel Synnaeve, Yossi Adi, and Alexandre D{\'e}fossez,
\newblock ``Simple and controllable music generation,''
\newblock {\em NeurIPS}, vol. 36, 2024.

\bibitem{valle}
Chengyi Wang, Sanyuan Chen, Yu~Wu, Ziqiang Zhang, Long Zhou, Shujie Liu, Zhuo Chen, Yanqing Liu, Huaming Wang, Jinyu Li, Lei He, Sheng Zhao, and Furu Wei,
\newblock ``Neural codec language models are zero-shot text to speech synthesizers,'' 2023.

\bibitem{audiobox}
Apoorv Vyas, Bowen Shi, Matthew Le, Andros Tjandra, Yi-Chiao Wu, Baishan Guo, Jiemin Zhang, Xinyue Zhang, Robert Adkins, William Ngan, Jeff Wang, Ivan Cruz, Bapi Akula, Akinniyi Akinyemi, Brian Ellis, Rashel Moritz, Yael Yungster, Alice Rakotoarison, Liang Tan, Chris Summers, Carleigh Wood, Joshua Lane, Mary Williamson, and Wei-Ning Hsu,
\newblock ``Audiobox: Unified audio generation with natural language prompts,'' 2023.

\bibitem{audiolm}
Zalán Borsos, Raphaël Marinier, Damien Vincent, Eugene Kharitonov, Olivier Pietquin, Matt Sharifi, Dominik Roblek, Olivier Teboul, David Grangier, Marco Tagliasacchi, and Neil Zeghidour,
\newblock ``Audiolm: a language modeling approach to audio generation,'' 2023.

\bibitem{stablediff}
Robin Rombach, Andreas Blattmann, Dominik Lorenz, Patrick Esser, and Björn Ommer,
\newblock ``High-resolution image synthesis with latent diffusion models,'' 2021.

\bibitem{gpt4}
OpenAI,
\newblock ``Gpt-4 technical report,'' 2023.

\bibitem{audioseal}
Robin~San Roman, Pierre Fernandez, Alexandre Défossez, Teddy Furon, Tuan Tran, and Hady Elsahar,
\newblock ``Proactive detection of voice cloning with localized watermarking,'' 2024.

\bibitem{wavmark}
Guangyu Chen, Yu~Wu, Shujie Liu, Tao Liu, Xiaoyong Du, and Furu Wei,
\newblock ``Wavmark: Watermarking for audio generation,'' 2024.

\bibitem{communication2023seamless}
Lo{\"\i}c Barrault, Yu-An Chung, Mariano~Coria Meglioli, David Dale, Ning Dong, Mark Duppenthaler, Paul-Ambroise Duquenne, Brian Ellis, Hady Elsahar, Justin Haaheim, et~al.,
\newblock ``Seamless: Multilingual expressive and streaming speech translation,''
\newblock {\em arXiv preprint arXiv:2312.05187}, 2023.

\bibitem{stablesignature}
Pierre Fernandez, Guillaume Couairon, Hervé Jégou, Matthijs Douze, and Teddy Furon,
\newblock ``The stable signature: Rooting watermarks in latent diffusion models,'' 2023.

\bibitem{kim2023wouaf}
Changhoon Kim, Kyle Min, Maitreya Patel, Sheng Cheng, and Yezhou Yang,
\newblock ``Wouaf: Weight modulation for user attribution and fingerprinting in text-to-image diffusion models,''
\newblock {\em arXiv preprint arXiv:2306.04744}, 2023.

\bibitem{kumar2019melgan}
Kundan Kumar, Rithesh Kumar, Thibault de~Boissiere, Lucas Gestin, Wei~Zhen Teoh, Jose Sotelo, Alexandre de~Brebisson, Yoshua Bengio, and Aaron Courville,
\newblock ``Melgan: Generative adversarial networks for conditional waveform synthesis,'' 2019.

\bibitem{kong2020hifi}
Jungil Kong, Jaehyeon Kim, and Jaekyoung Bae,
\newblock ``Hifi-gan: Generative adversarial networks for efficient and high fidelity speech synthesis,'' 2020.

\bibitem{mbd}
Robin~San Roman, Yossi Adi, Antoine Deleforge, Romain Serizel, Gabriel Synnaeve, and Alexandre Défossez,
\newblock ``From discrete tokens to high-fidelity audio using multi-band diffusion,''
\newblock {\em NeurIPS}, vol. 36, 2024.

\bibitem{encodec}
Alexandre Défossez, Jade Copet, Gabriel Synnaeve, and Yossi Adi,
\newblock ``High fidelity neural audio compression,''
\newblock {\em arXiv preprint arXiv:2210.13438}, 2022.

\bibitem{oord2016wavenet}
Aaron van~den Oord, Sander Dieleman, Heiga Zen, Karen Simonyan, Oriol Vinyals, Alex Graves, Nal Kalchbrenner, Andrew Senior, and Koray Kavukcuoglu,
\newblock ``Wavenet: A generative model for raw audio,''
\newblock {\em arXiv preprint arXiv:1609.03499}, 2016.

\bibitem{radford2018improving}
Alec Radford, Karthik Narasimhan, Tim Salimans, Ilya Sutskever, et~al.,
\newblock ``Improving language understanding by generative pre-training,''
\newblock 2018.

\bibitem{brown2020language}
Tom Brown, Benjamin Mann, Nick Ryder, Melanie Subbiah, Jared~D Kaplan, Prafulla Dhariwal, Arvind Neelakantan, Pranav Shyam, Girish Sastry, Amanda Askell, et~al.,
\newblock ``Language models are few-shot learners,''
\newblock {\em NeurIPS}, vol. 33, pp. 1877--1901, 2020.

\bibitem{kharitonov2021text}
Eugene Kharitonov, Ann Lee, Adam Polyak, Yossi Adi, Jade Copet, Kushal Lakhotia, Tu-Anh Nguyen, Morgane Rivi{\`e}re, Abdelrahman Mohamed, Emmanuel Dupoux, et~al.,
\newblock ``Text-free prosody-aware generative spoken language modeling,''
\newblock {\em arXiv preprint arXiv:2109.03264}, 2021.

\bibitem{soundstream}
Neil Zeghidour, Alejandro Luebs, Ahmed Omran, Jan Skoglund, and Marco Tagliasacchi,
\newblock ``Soundstream: An end-to-end neural audio codec,'' 2021.

\bibitem{dac}
Rithesh Kumar, Prem Seetharaman, Alejandro Luebs, Ishaan Kumar, and Kundan Kumar,
\newblock ``High-fidelity audio compression with improved rvqgan,'' 2023.

\bibitem{vaswani2017attention}
Ashish Vaswani, Noam Shazeer, Niki Parmar, Jakob Uszkoreit, Llion Jones, Aidan~N Gomez, {\L}ukasz Kaiser, and Illia Polosukhin,
\newblock ``Attention is all you need,''
\newblock {\em NeurIPS}, vol. 30, 2017.

\bibitem{voicebox}
Matthew Le, Apoorv Vyas, Bowen Shi, Brian Karrer, Leda Sari, Rashel Moritz, Mary Williamson, Vimal Manohar, Yossi Adi, Jay Mahadeokar, and Wei-Ning Hsu,
\newblock ``Voicebox: Text-guided multilingual universal speech generation at scale,'' 2023.

\bibitem{naturalspeech2}
Kai Shen, Zeqian Ju, Xu~Tan, Yanqing Liu, Yichong Leng, Lei He, Tao Qin, Sheng Zhao, and Jiang Bian,
\newblock ``Naturalspeech 2: Latent diffusion models are natural and zero-shot speech and singing synthesizers,'' 2023.

\bibitem{jasco}
Or~Tal, Alon Ziv, Itai Gat, Felix Kreuk, and Yossi Adi,
\newblock ``Joint audio and symbolic conditioning for temporally controlled text-to-music generation,'' 2024.

\bibitem{stableaudio}
Zach Evans, CJ~Carr, Josiah Taylor, Scott~H. Hawley, and Jordi Pons,
\newblock ``Fast timing-conditioned latent audio diffusion,'' 2024.

\bibitem{lie2006robust}
Wen-Nung Lie and Li-Chun Chang,
\newblock ``Robust and high-quality time-domain audio watermarking based on low-frequency amplitude modification,''
\newblock {\em IEEE transactions on multimedia}, vol. 8, no. 1, pp. 46--59, 2006.

\bibitem{kalantari2009robust}
Nima~Khademi Kalantari, Mohammad~Ali Akhaee, Seyed~Mohammad Ahadi, and Hamidreza Amindavar,
\newblock ``Robust multiplicative patchwork method for audio watermarking,''
\newblock {\em IEEE Transactions on Audio, speech, and language processing}, vol. 17, no. 6, pp. 1133--1141, 2009.

\bibitem{liu2023dear}
Chang Liu, Jie Zhang, Han Fang, Zehua Ma, Weiming Zhang, and Nenghai Yu,
\newblock ``Dear: A deep-learning-based audio re-recording resilient watermarking,''
\newblock in {\em Proceedings of the AAAI Conference on Artificial Intelligence}, 2023, vol.~37, pp. 13201--13209.

\bibitem{pavlovic2022robust}
Kosta Pavlovi{\'c}, Slavko Kova{\v{c}}evi{\'c}, Igor Djurovi{\'c}, and Adam Wojciechowski,
\newblock ``Robust speech watermarking by a jointly trained embedder and detector using a dnn,''
\newblock {\em Digital Signal Processing}, vol. 122, pp. 103381, 2022.

\bibitem{o2024maskmark}
Patrick O’Reilly, Zeyu Jin, Jiaqi Su, and Bryan Pardo,
\newblock ``Maskmark: Robust neuralwatermarking for real and synthetic speech,''
\newblock in {\em ICASSP}. IEEE, 2024, pp. 4650--4654.

\bibitem{yu2021responsible}
Ning Yu, Vladislav Skripniuk, Dingfan Chen, Larry~S Davis, and Mario Fritz,
\newblock ``Responsible disclosure of generative models using scalable fingerprinting,''
\newblock in {\em International Conference on Learning Representations}, 2021.

\bibitem{juvela2023collaborative}
Lauri Juvela and Xin Wang,
\newblock ``Collaborative watermarking for adversarial speech synthesis,''
\newblock {\em arXiv preprint arXiv:2309.15224}, 2023.

\bibitem{yu2021artificial}
Ning Yu, Vladislav Skripniuk, Sahar Abdelnabi, and Mario Fritz,
\newblock ``Artificial fingerprinting for generative models: Rooting deepfake attribution in training data,''
\newblock in {\em Proceedings of the IEEE/CVF International conference on computer vision}, 2021, pp. 14448--14457.

\bibitem{zhao2023recipe}
Yunqing Zhao, Tianyu Pang, Chao Du, Xiao Yang, Ngai-Man Cheung, and Min Lin,
\newblock ``A recipe for watermarking diffusion models,''
\newblock {\em arXiv preprint arXiv:2303.10137}, 2023.

\bibitem{gu2023learnability}
Chenchen Gu, Xiang~Lisa Li, Percy Liang, and Tatsunori Hashimoto,
\newblock ``On the learnability of watermarks for language models,''
\newblock {\em arXiv preprint arXiv:2312.04469}, 2023.

\bibitem{sander2024watermarking}
Tom Sander, Pierre Fernandez, Alain Durmus, Matthijs Douze, and Teddy Furon,
\newblock ``Watermarking makes language models radioactive,''
\newblock {\em arXiv preprint arXiv:2402.14904}, 2024.

\bibitem{touvron2023llama}
Hugo Touvron, Louis Martin, Kevin Stone, Peter Albert, Amjad Almahairi, Sharan Narang, Aurelien Rodriguez, Robert Stojnic, Sergey Edunov, Thomas Scialom, et~al.,
\newblock ``Llama 2: Open foundation and fine-tuned chat models,'' 2023.

\bibitem{kong2021diffwave}
Zhifeng Kong, Wei Ping, Jiaji Huang, Kexin Zhao, and Bryan Catanzaro,
\newblock ``Diffwave: A versatile diffusion model for audio synthesis,'' 2021.

\bibitem{fad}
Kevin Kilgour, Mauricio Zuluaga, Dominik Roblek, and Matthew Sharifi,
\newblock ``Fr$\backslash$'echet audio distance: A metric for evaluating music enhancement algorithms,''
\newblock {\em arXiv preprint arXiv:1812.08466}, 2018.

\end{thebibliography}
\end{document}